\documentclass{article}

\usepackage{microtype}
\usepackage{graphicx}
\usepackage{subfigure}
\usepackage{booktabs} 

\usepackage{hyperref}

\usepackage[accepted]{icml2025}

\usepackage{amsmath}
\usepackage{amssymb}
\usepackage{mathtools}
\usepackage{amsthm}
\usepackage{subcaption}

\usepackage[capitalize,noabbrev]{cleveref}

\theoremstyle{plain}

\theoremstyle{definition}

\theoremstyle{remark}

\usepackage[normalem]{ulem}
\usepackage{color}
\usepackage{xcolor}
\usepackage{enumitem}
\usepackage{multirow}

\usepackage[textsize=tiny]{todonotes}

\icmltitlerunning{TransDiffSBDD: Causality-Aware Multi-Modal Structure-Based Drug Design}

\begin{document}

\twocolumn[
\icmltitle{TransDiffSBDD: Causality-Aware Multi-Modal Structure-Based Drug Design}



\icmlsetsymbol{equal}{*}

\begin{icmlauthorlist}
\icmlauthor{Xiuyuan Hu}{thuee}
\icmlauthor{Guoqing Liu}{msrai4s}
\icmlauthor{Can (Sam) Chen}{mcgill,mila}
\icmlauthor{Yang Zhao}{thuee}
\icmlauthor{Hao Zhang$^\ast$}{thuee}
\icmlauthor{Xue Liu}{mcgill,mila}
\end{icmlauthorlist}

\icmlaffiliation{thuee}{Tsinghua University}
\icmlaffiliation{msrai4s}{Microsoft Research AI for Science}
\icmlaffiliation{mcgill}{McGill University}
\icmlaffiliation{mila}{Mila - Quebec AI Institute}

\icmlcorrespondingauthor{Hao Zhang}{haozhang@tsinghua.edu.cn}

\icmlkeywords{Machine Learning, ICML}

\vskip 0.3in 
]



\printAffiliationsAndNotice{} 

\begin{abstract}
Structure-based drug design (SBDD) is a critical task in drug discovery, requiring the generation of molecular information across two distinct modalities: discrete molecular graphs and continuous 3D coordinates. 
However, existing SBDD methods often overlook two key challenges: (1) the multi-modal nature of this task and (2) the causal relationship between these modalities, limiting their plausibility and performance. 
To address both challenges, we propose \textbf{TransDiffSBDD}, an integrated framework combining autoregressive transformers and diffusion models for SBDD. Specifically, the autoregressive transformer models discrete molecular information, while the diffusion model samples continuous distributions, effectively resolving the first challenge. To address the second challenge, we design a hybrid-modal sequence for protein-ligand complexes that explicitly respects the causality between modalities. Experiments on the CrossDocked2020 benchmark demonstrate that TransDiffSBDD outperforms existing baselines.
\end{abstract}

\section{Introduction}
Artificial intelligence (AI) has recently achieved remarkable success in solving various scientific discovery tasks, including protein structure prediction \cite{AlphaFold2} and weather forecasting \cite{WeatherForcastingHuawei}. Among these, AI-driven drug discovery (AIDD) has emerged as a highly promising area \cite{AIDDsurvey}. Traditional drug development processes are notoriously time-consuming and resource-intensive, and machine learning (ML) techniques have the potential to alleviate these challenges \cite{MLforDDsurvey} by significantly improving the accuracy and efficiency of tasks such as molecular property prediction \cite{PropPredSurvey} and retrosynthesis \cite{RetrosynSurvey}. Among the mainstream strategies in pharmaceutical research, structure-based drug design stands out as a pivotal challenge in AIDD. SBDD is a 3D molecular structure generation task that aims to design drug-like and synthesizable small-molecule ligands with high binding affinity for a given protein target \cite{SBDDsurvey}.

Despite the development of numerous ML methods for SBDD in recent years, their performance remains unsatisfactory \cite{SBDDsurvey2}. This limitation primarily arises from the inadequate consideration of the relationship between the two modalities in molecular information: the discrete graph information (atom types and chemical bonds) and the continuous 3D information (atomic coordinates). Two key factors contribute to this limitation: (1) the technical discrepancies in modeling the two modalities, and (2) the causal relationship between them.

First, the outputs of SBDD involve both discrete and continuous modalities, making SBDD a prototypical multi-modal generation task. Discrete information, including atom types and bonds, is well-suited to autoregressive language models based on transformer architectures, which have revolutionized natural language processing (NLP) \cite{GPT-3,InstructGPT,Llama}. Conversely, continuous information of 3D atomic coordinates aligns with diffusion-based generative models that have demonstrated outstanding performance in image and video generation \cite{LatentDiffusion,StableVideoDiffusion}. So far, while many diffusion-based models for SBDD have been proposed (e.g., \citet{TargetDiff}), they often struggle to effectively capture discrete molecular features. Similarly, some autoregressive language models have been adapted for SBDD (e.g., \citet{BindGPT}), but they often represent 3D coordinates as discrete tokens, compromising their ability to model continuous distributions. As a result, neither architecture alone is well-suited for the multi-modal nature of SBDD. An effective solution requires combining these two architectures to better align with the intrinsic demands of SBDD.

Second, there is a causal relationship between a molecule's discrete graph information and its continuous 3D structure in the context of a protein binding pocket. Specifically, once the graph structure of a ligand is determined, its 3D binding pose is largely dictated\footnote{Environmental factors such as the temperature may have an impact, so we use "largely" rather than "completely."}, and protein-ligand docking aims to model this causality \cite{AutoDockVina,DockingSurvey}. However, most existing 3D generation methods for SBDD generate discrete and continuous molecular information simultaneously (e.g., \citet{Pocket2Mol,DecompDiff}), neglecting this causality. Benchmark results in \citet{DrugDesignChallenge} demonstrate that combining 1D/2D molecular generation methods with docking software can outperform 3D methods specialized for SBDD. Although 1D/2D methods do not explicitly utilize structural information from protein targets, their integration with docking software respects the causality between discrete and continuous molecular information. This explains their superior performance and highlights a critical flaw in current 3D approaches: their neglect of this causality.

To address these challenges, we propose \textbf{TransDiffSBDD}, a framework for structure-based drug design integrating autoregressive transformer and diffusion architectures. TransDiffSBDD not only accommodates the multi-modal nature of SBDD by integrating the two architectures but also explicitly models the causality between discrete 2D and continuous 3D molecular information. 

Specifically, we design a hybrid-modal sequence to represent protein-ligand complexes, where molecular discrete graph information is encoded by SMILES \cite{SMILES} strings, and atomic 3D coordinates are represented as numerical vectors. This sequence format explicitly preserves the causality (the 2nd challenge) by placing all the 3D coordinates after SMILES tokens, and modeling the SBDD problem as an autoregressive generation task. Moreover, inspired by recent advances in multi-modal generative models \cite{Transfusion,LatentLM}, we introduce the first integrated architecture of an autoregressive transformer and a diffusion for SBDD. The transformer model provides a global understanding of the hybrid-modal sequence and generates molecular discrete graph information, while the diffusion component focuses on modeling the continuous distribution of atomic 3D coordinates. Therefore, the multi-modal dilemma (the 1st challenge) of SBDD is effectively solved by the respective roles of the two architectures. Additionally, we utilize a joint loss function for the integrated model, which combines cross-entropy loss for discrete tokens and diffusion loss for continuous vectors. During training, we first perform pretraining on a large-scale small molecule dataset and a protein-ligand complex dataset, and then apply reinforcement learning (RL)-based finetuning for each protein target, yielding a set of target-specific 3D molecular candidates.

Experimental results on the CrossDocked2020 benchmark \cite{CrossDocked2020} demonstrate that TransDiffSBDD outperforms existing state-of-the-art SBDD baselines, including autoregressive models, diffusion-based methods, and approaches that combine 1D/2D generation with docking. Notably, TransDiffSBDD achieves an outstanding Success Rate in drug design towards multi-property objectives (MPO), reflecting its superior performance in real-world SBDD. Case studies further validate the practical effectiveness of our method. Moreover, our framework has the potential to seamlessly adapt to datasets containing distributional information of molecular 3D structures, which represents an important direction for the future of AIDD.

In summary, our main contributions are as follows:
\begin{itemize}
    \item We raise and address the primary limitations of existing SBDD methods: the neglect of the modal discrepancy and causality between molecular discrete graph information and continuous 3D information.
    \item We propose TransDiffSBDD, a novel framework for structure-based drug design that integrates autoregressive transformer and diffusion architectures to generate hybrid-modal sequences for protein-ligand complexes.
    \item TransDiffSBDD surpasses multiple baselines on CrossDocked2020, demonstrating its robustness and effectiveness in structure-based drug design.
\end{itemize}
\section{Related Works}
\subsection{Structure-based Drug Design}
Structure-based drug design is a critical 3D molecular generation problem in pharmacology, where protein pockets with biomedical significance serve as explicit targets for designing small-molecule ligands \cite{SBDDsurvey3}. Recent SBDD approaches can be broadly categorized into two main paradigms:
\begin{itemize}
    \item Discrete-focused modeling, which primarily relies on autoregressive models, including AR \cite{AR}, Pocket2Mol \cite{Pocket2Mol}, Lingo3DMol \cite{Lingo3DMol}, FLAG \cite{FLAG}, ResGen \cite{ResGen}, XYZ-TF \cite{XYZ-TF}, BindGPT \cite{BindGPT}, and 3DMolFormer \cite{3DMolFormer}.
    \item Continuous-focused modeling, which typically employs denoising models such as diffusion and flow networks, including TargetDiff \cite{TargetDiff}, DecompDiff \cite{DecompDiff}, DrugGPS \cite{DrugGPS}, PocketFlow \cite{PocketFlow}, IPDiff \cite{IPDiff}, IRDiff \cite{IRDiff}, AliDiff \cite{AliDiff}, D3FG \cite{D3FG}, DiffSBDD \cite{DiffSBDD}, MolCRAFT \cite{MolCRAFT}, FlexSBDD \cite{FlexSBDD}, and VoxBind \cite{VoxBind}.
\end{itemize}

Unfortunately, none of the above methods can balance the modeling of discrete and continuous modes. Additionally, \citet{DrugDesignChallenge} highlights the competitiveness of 1D/2D molecular generation methods combined with docking software in SBDD. Among 1D/2D approaches, reinforcement learning has emerged as a dominant technique, where the property objective serves as the RL reward. Representative methods in this category include Reinvent \cite{Reinvent,Reinvent4}, RationaleRL \cite{RationaleRL}, RGA\cite{RGA}, and ChemRLFormer \cite{ChemRLFormer}.

\subsection{Multi-modal Generative Models}
Multi-modal large language models (MLLMs) have become one of the most actively researched topics, aiming to develop unified architectures capable of understanding and generating data across multiple modalities, such as text, images, audio, and video \cite{MLLMsurvey}. A central challenge in this area is handling the inherent discrepancy among these modalities. Several approaches have been proposed for multi-modal generation, with a recently popular paradigm being the integration of autoregressive transformers and diffusion models \cite{Transfusion}. This approach restructures multi-modal data into a hybrid-modal sequence, which is then processed autoregressively by a transformer backbone, while the diffusion model serves as an output head for sampling continuous data (e.g., images). This strategy has not only demonstrated effectiveness in typical multi-modal scenarios \cite{LatentLM,LlamaFusion} but also shows promising potential for applications in other fields, such as embodied AI \cite{Embodied} and general scientific tasks \cite{UniGenX}.
\section{Preliminaries}

\subsection{Autoregressive Transformers for Sequence Modeling}
Autoregressive transformers are a class of neural networks widely used for sequence generation tasks. Their architecture is based on the transformer model, which was introduced to efficiently handle sequential data using self-attention mechanisms \cite{Transformer}. In autoregressive modeling, the transformer generates sequences element by element in a causal manner, ensuring that the prediction of each element depends only on previous elements in the sequence. Formally, for an ordered sequence $\{x^1,x^2,...,x^n\}$, autoregressive generation is formulated as:
\begin{equation}
p(x^1,x^2,...,x^n)=\prod_{i=1}^n p(x^i|x^1,...,x^{i-1}),
\end{equation}
and in autoregressive transformers the conditional probability is achieved using masked self-attention, where the attention mechanism is constrained to prevent access to future elements during training and inference. Representative architectures such as GPT (Generative Pretrained Transformer) \cite{GPT-2} have demonstrated remarkable success in tasks ranging from natural language generation \cite{GPT-3} to molecular design \cite{MolGPT}, where sequences like SMILES strings are modeled as text.

\subsection{Diffusion Models}
Diffusion models, also known as denoising diffusion probabilistic models (DDPMs) \cite{DDPM}, have emerged as powerful generative models for continuous data, such as images \cite{LatentDiffusion}, videos \cite{StableVideoDiffusion}, and molecular 3D poses \cite{DiffDock}. DDPMs reverse a diffusion process, which gradually corrupts data into noise, to reconstruct the original data.

The forward diffusion process iteratively adds Gaussian noise to the data in a fixed number of steps $T$, transforming the data into pure noise at the final step. Formally, given data $x_0$, the noisy version at step $t$, denoted as $x_t$, is obtained by:
\begin{equation}
x_t = \sqrt{\alpha_t} x_0 + \sqrt{1 - \alpha_t} \epsilon, \quad \epsilon \sim \mathcal{N}(0, I),
\end{equation}
where $\alpha_t$ is a variance schedule controlling the noise level at each step.

In the reverse diffusion process, a neural network parameterized by $\theta$ is trained to denoise $x_t$ and predict $x_0$. This is achieved by minimizing a loss function that measures the mean squared error (MSE) between the predicted noise and the actual noise $\epsilon$ added during the forward process:
\begin{equation}
\label{DDPMloss}
\mathcal{L}_{\text{DDPM}}(z,x)=\mathbb{E}_{\epsilon,t}\left[ \lVert\epsilon-\epsilon_\theta(x_t,t,z)\rVert^2\right],
\end{equation}
where the denoising network $\epsilon_\theta$ predicts the noise by $x_t$, $t$, and the conditional information $z$.

\subsection{Diffusion Loss for Autoregressive Models}
To implement autoregressive image generation, \citet{MAR} introduced the diffusion loss for autoregressive models, where the conditional vector $z^i$ is generated by the autoregressive network $f(\cdot)$ operating on previous elements: $z^i=f(x^1,...,x^{i-1})$. In this way, the diffusion loss in Eqn. (\ref{DDPMloss}) can be applied to autoregressive models:
\begin{equation}
\label{DDPMARloss}
\mathcal{L}_{\text{DDPM-AR}}(x)=\mathbb{E}_{x^i,\epsilon,t}\left[\lVert \epsilon-\epsilon_\theta(x_t^i,t,f(x^1,...,x^{i-1}))\rVert^2\right],
\end{equation}
where the diffusion parameters $\theta$ and the parameters of the autoregressive model $f(\cdot)$ are trained together.
\section{TransDiffSBDD}
\begin{figure*}[t]
    \centering
    \begin{subfigure}
        \centering
        \includegraphics[width=0.95\linewidth]{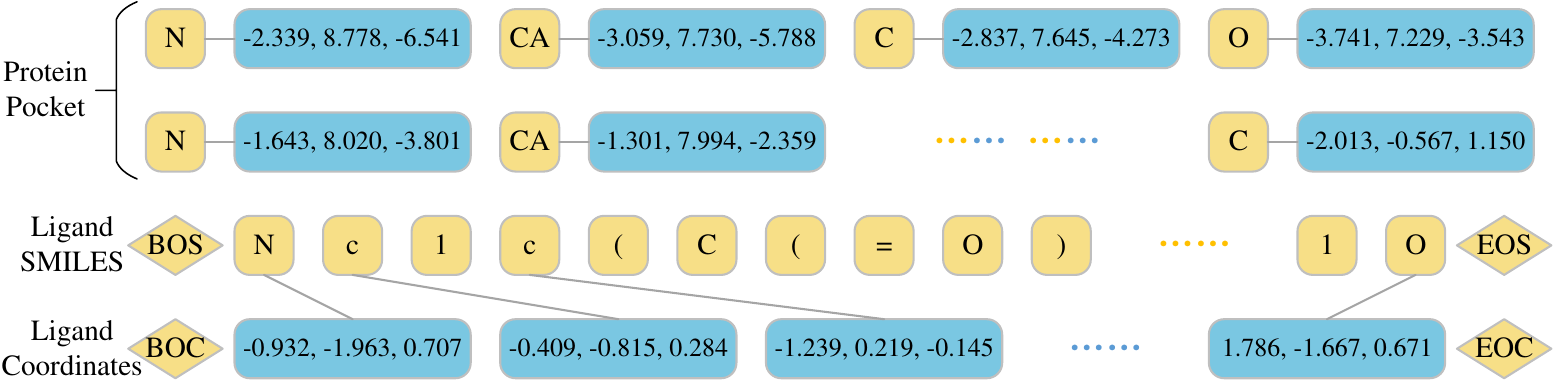}
        \caption*{(a) Hybrid-modal sequence for protein-ligand complexes.}
    \end{subfigure}
    \vspace{1em}
    \begin{subfigure}
        \centering
        \includegraphics[width=0.95\linewidth]{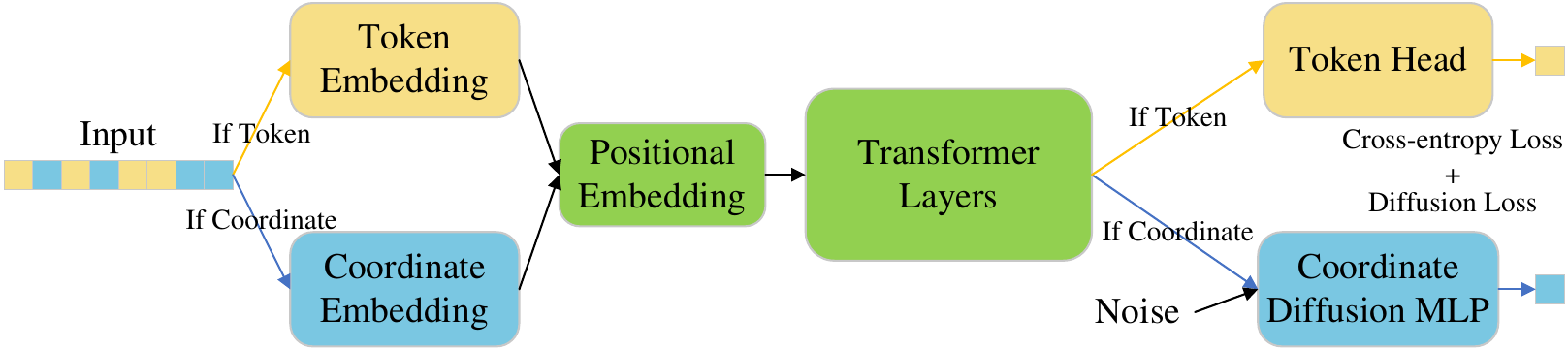}
        \caption*{(b) Integrated model of autoregressive transformer and diffusion.}
    \end{subfigure}
    \caption{Overview of the TransDiffSBDD framework. (a) This figure illustrates the hybrid-modal sequence for a protein-ligand complex, where discrete tokens are marked in yellow, continuous 3D coordinates are marked in blue, and the connections indicate the correspondence between atoms and their coordinates. The sequence consists of alternating atomic and coordinate information for the protein structure, followed by the ligand's discrete graph information represented as SMILES and its 3D coordinate information. (b) This diagram depicts the integrated model of autoregressive transformer and diffusion, where components dedicated to discrete tokens are marked in yellow, components specialized for continuous 3D coordinates are marked in blue, and shared modeling components are marked in green. Specifically, when the output is 3D coordinates, the output vector from the transformer layers serves as conditional information for the diffusion MLP. }
    \label{overview}
\end{figure*}

This section details the TransDiffSBDD framework for structure-based drug design (SBDD), encompassing: (1) hybrid-modal sequences for protein-ligand complexes, which separate the two modalities of SBDD information and respect the causality between them. (2) an integrated model of autoregressive transformers and diffusion, where the transformer and diffusion components are applied to discrete and continuous modalities, respectively. (3) training techniques employed in TransDiffSBDD, including a joint loss function for two modalities, reinforcement learning for specific protein targets, and preprocessing and augmentation of training data.

\subsection{Hybrid-modal Sequences for Protein-ligand Complexes}
To separately handle the discrete and continuous modalities in SBDD, we design a hybrid-modal sequence format to represent protein-ligand complexes, consisting of discrete tokens and 3D coordinates (represented as 3-dimensional numerical vectors), as illustrated in Figure \ref{overview}(a). The sequence includes the structural information of both the protein pocket and the ligand. The ligand's atomic coordinates follow the discrete graph information encoded as a SMILES string, preserving the causality between the molecular information of the two modalities in the SBDD task.

Specifically, the protein pocket information is directly extracted from PDB-formatted data, with each atom represented by its discrete token followed by its 3D coordinates. Hydrogen atoms in the protein pocket are removed, while other atoms are denoted by their types, such as 'N,' 'C,' 'O,' and 'S.' The alpha carbon in each amino acid is specially labeled as 'CA,' and each amino acid starts with the ['N,' 'CA,' 'C,' 'O'] sequence, consistent with the PDB format.

For the ligand, the discrete graph information includes atom types and chemical bonds, which are represented using SMILES, a widely adopted format for molecular generation. The SMILES strings are tokenized at the atomic level \cite{SMILEStokenization}, with each token corresponding to an atom, a number, or a symbol. The ligand's 3D coordinate sequence aligns sequentially with the atomic tokens in the SMILES string, while numbers and symbols lack corresponding coordinates. The beginning and end of the SMILES sequence are denoted by 'BOS' and 'EOS,' while the coordinate sequence is marked by 'BOC' and 'EOC.'

Using a causal autoregressive model to generate these hybrid-modal sequences provides an intuitive solution for SBDD, where the protein pocket subsequence serves as input, and the ligand SMILES and 3D coordinates are autoregressively sampled during inference, ensuring the causality between the two modalities.

\subsection{Integrated Model of Autoregressive Transformer and Diffusion}
We have formulated the SBDD task as a sequence generation problem, where causal autoregressive transformer models, such as GPT, are well-suited for such scenarios. However, original GPT architectures are limited to discrete token modeling and cannot directly sample from continuous distributions. This challenge also arises in general multi-modal generation tasks, which aim to unify the processing of discrete data (e.g., text) and continuous data (e.g., images, audio, and video).

A promising recent solution for multi-modal generation involves tokenizing continuous data (e.g., splitting images into patches) and constructing sequences with discrete data \cite{Transfusion,LatentLM,LlamaFusion}. An autoregressive transformer serves as the backbone for embedding the sequence data, while a diffusion model acts as the output head to sample continuous distributions. We observe that the hybrid-modal sequence designed for SBDD aligns well with this architecture.

Consequently, we develop an integrated model of autoregressive transformers and diffusion, inspired by advancements in multi-modal generation, as illustrated in Figure \ref{overview}(b). Specifically, we adopt a GPT-like \cite{GPT-2} backbone comprising embedding layers, causal transformer layers, and output layers. The embedding layers include a token embedding block for discrete tokens and a linear embedding layer for mapping 3D coordinates to the same dimensional space as token embeddings, followed by a positional embedding block. The output layers include an MLP token head, similar to original GPT models, and a denoising diffusion network implemented with an MLP to output 3D coordinates. The outputs of the transformer layers serve as conditioning inputs for the diffusion model.

The diffusion model conducts a denoising process that generates a ligand's 3D coordinates from noise, conditioned on the structural information of the protein pocket and the ligand's discrete graph. During inference, only the ligand's atomic coordinates (typically fewer than 30) need to be generated. Compared to image patches, the lower dimensionality of 3D coordinates reduces the computational complexity of the MLP network and the number of denoising steps, making inference computationally lightweight. Additionally, during inference, the structural information of the protein pocket is used as input. The output layer is first repeatedly called as the token head to generate discrete tokens until the 'EOS' token is generated, marking the end of the SMILES sequence. The next token is then forced to be 'BOC,' after which the diffusion model is invoked to generate the 3D coordinates of the ligand atoms. The diffusion process is called as many times as there are atoms in the SMILES sequence. This design minimizes the computational complexity during the inference phase.

Notably, the diffusion model introduces stochasticity through noise, which facilitates the modeling of atomic coordinate distributions. Since atomic spatial positions inherently follow an equilibrium distribution (e.g., multiple binding modes in protein-ligand docking), this stochasticity is advantageous. Although current datasets are insufficient for training, our approach provides a potential solution for modeling molecular equilibrium distributions, which is a significant future direction \cite{DiG}.

\subsection{Training Techniques}
\begin{table*}[tb]
\centering
\caption{Experimental results of TransDiffSBDD and other baselines on CrossDocked2020. Some results are from those reported in DecompDiff, DecompOpt, and MolCRAFT. ($\uparrow$) / ($\downarrow$) indicates that a higher / lower value is better. The 1st and 2nd best results in each column are \textbf{bolded} and \underline{underlined}, respectively.}
\label{results}
\begin{tabular}{cccccccc}
\hline
Methods & Vina Score ($\downarrow$) & Vina Dock ($\downarrow$) & QED ($\uparrow$) & SA ($\uparrow$) & Diversity ($\uparrow$) & Success Rate ($\uparrow$)
\\ \hline 
Reference Set & -6.36 & -7.45 & 0.48 & 0.73 & - & 25.0\% \\ \hline
Reinvent + Vina & - & -9.18 & 0.49 & 0.72 & \textbf{0.83} & \underline{76.7\%}\\
RGA + Vina & - & -8.01 & \textbf{0.57} & 0.71 & 0.41 & 46.2\% \\ \hline
AR & -5.75 & -6.75 & 0.51 & 0.63 & 0.70 & 7.1\% \\
liGAN & - & -6.33 & 0.39 & 0.59 & 0.66 & 3.9\% \\
GraphBP & - & -4.80 & 0.43 & 0.49 & 0.79 & 0.1\% \\
Pocket2Mol & -5.14 & -7.15 & \underline{0.56} & \underline{0.74} & 0.69 & 24.4\% \\
TargetDiff & -5.47 & -7.80 & 0.48 & 0.58 & 0.72 & 10.5\% \\
FLAG & - & -5.63 & 0.49 & 0.70 & 0.70 & 14.1\% \\
DecompDiff & -5.67 & -8.39 & 0.45 & 0.61 & 0.68 & 24.5\% \\
DecompOpt & -5.87 & -8.98 & 0.48 & 0.65 & 0.60 & 52.5\% \\
MolCRAFT & \textbf{-6.61} & \underline{-9.25} & 0.46 & 0.62 & 0.61 & 36.1\% \\
\hline
TransDiffSBDD & \underline{-6.02} & \textbf{-9.37} & 0.48 & \textbf{0.75} & \underline{0.81} & \textbf{83.9\%} \\ \hline
\end{tabular}
\end{table*}

\paragraph{Joint loss of two modalities}
In supervised training on protein-ligand complexes, the loss is calculated for the ligand part of each hybrid-modal sequence, encompassing discrete SMILES tokens and continuous 3D coordinates. These are computed by the token head and diffusion components, respectively. Inspired by the design in \citet{Transfusion}, we compute the cross-entropy (CE) loss for discrete tokens and the diffusion loss for 3D coordinates. The weighted sum of these losses forms the objective minimized during training:
\begin{equation}
\label{JointLoss}
\mathcal{L}_{\text{TransDiffSBDD}}=\mathcal{L}_{\text{CE}}+\lambda\cdot\mathcal{L}_{\text{DDPM-AR}},
\end{equation}
where $\lambda$ is a balancing coefficient. Particularly, the diffusion loss for autoregressive transformer optimizes both the transformer backbone network and the MLP noise estimator.

\paragraph{Reinforcement learning}
Reinforcement learning (RL) is commonly used in autoregressive molecular generation to optimize for specific property objectives and mitigate data scarcity for specific targets \cite{Reinvent,ChemRLFormer}. After pretraining the transformer backbone, we treat it as an RL agent while keeping the diffusion MLP parameters fixed. A specified molecular property evaluator $R(\cdot)$ serves as the RL reward function. During each RL step, the integrated model samples a set of 3D compounds, which are used to compute the regularized maximum likelihood estimate (MLE) loss to update the RL agent:
\begin{equation}
\label{RLloss}
\mathcal{L}_{RL}(x;\Theta)=(\log p_{\Theta_0}(x)+\mu\cdot R(x)-\log p_{\Theta}(x))^2,
\end{equation}
where $x$ is a ligand sampled by the RL agent, $\Theta$ refers to the parameters of the agent, $\Theta_0$ is their original values obtained by supervised training, and $\mu$ is a weighting coefficient. This loss function has proven simple yet effective for autoregressive molecular generation \cite{SMILES_RL}. Iterative RL optimization of the agent model can improve the expected score of the generated ligands on the specified properties, such as the binding affinity against a certain protein target.

\paragraph{Data preprocessing and augmentation}
In preparing protein-ligand complex data for training, we remove samples containing rare elements like metals in ligands and apply ligand SMILES randomization \cite{SMILESRandomization}, a common data augmentation technique for molecular data. The order of atomic coordinates in the hybrid-modal sequence changes with the SMILES string's atom order. Moreover, we translate and rotate the 3D coordinates of protein pockets and ligands, normalizing the ligand's center of mass at the origin and applying random 3D rotations to reduce overfitting in scenarios with limited 3D data:
\begin{equation}
\left[\begin{array}{ccc}
x_1', ..., x_n' \\
y_1', ..., y_n' \\
z_1', ..., z_n' \\
\end{array}\right]=R^3\cdot
\left[\begin{array}{ccc}
x_1-x_c, ..., x_n-x_c \\
y_1-x_c, ..., y_n-x_c \\
z_1-x_c, ..., z_n-x_c \\
\end{array}\right],
\end{equation}
where $(x_i,y_i,z_i)$ is the original 3D coordinates of the $i$-th atom in a protein-ligand complex, $(x_c,y_c,z_c)$ represents the position of the ligand's center of mass, $R^3$ is a random 3D rotation matrix shaped $3\times3$, and $(x_i',y_i',z_i')$ refers to the coordinates for model training.
\section{Experiments}
\begin{figure*}[tb]
\centering
\begin{minipage}{0.19\textwidth}
    \centering
    \includegraphics[width=\textwidth]{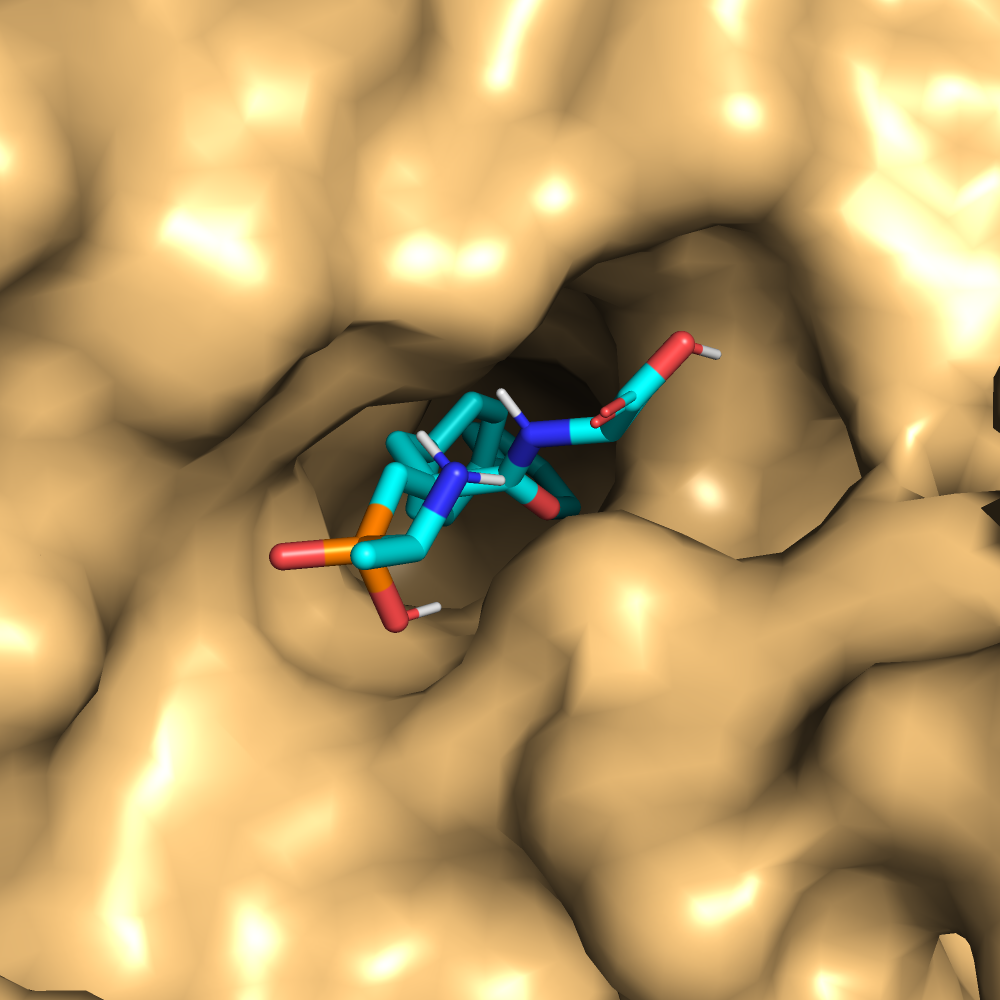} \\ \vspace{0.4cm}
    \includegraphics[width=\textwidth]{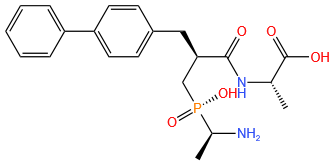}
    \centerline{Reference}
\end{minipage}
\begin{minipage}{0.19\textwidth}
    \centering
    \includegraphics[width=\textwidth]{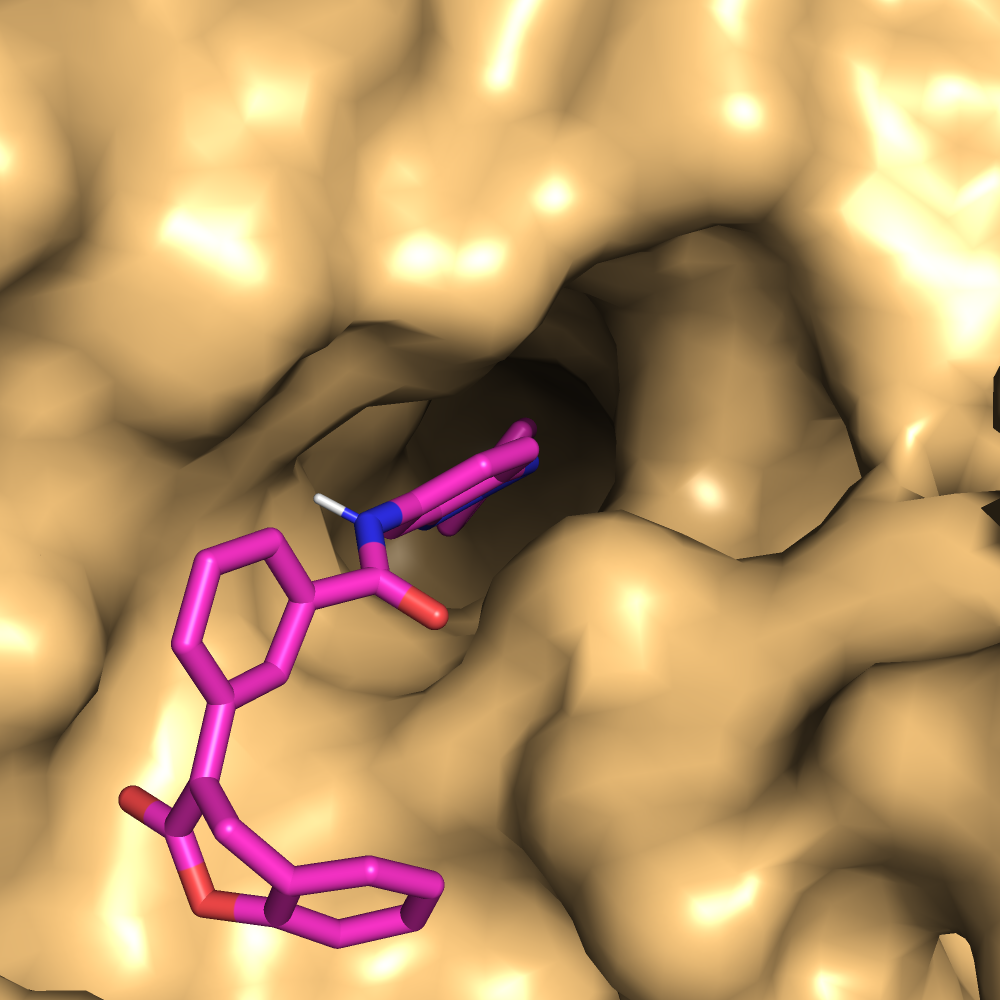} \\ \vspace{0.4cm}
    \includegraphics[width=\textwidth]{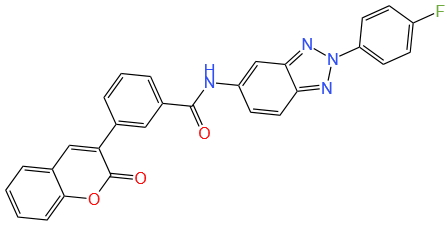}
    \centerline{Candidate 1}
\end{minipage}
\begin{minipage}{0.19\textwidth}
    \centering
    \includegraphics[width=\textwidth]{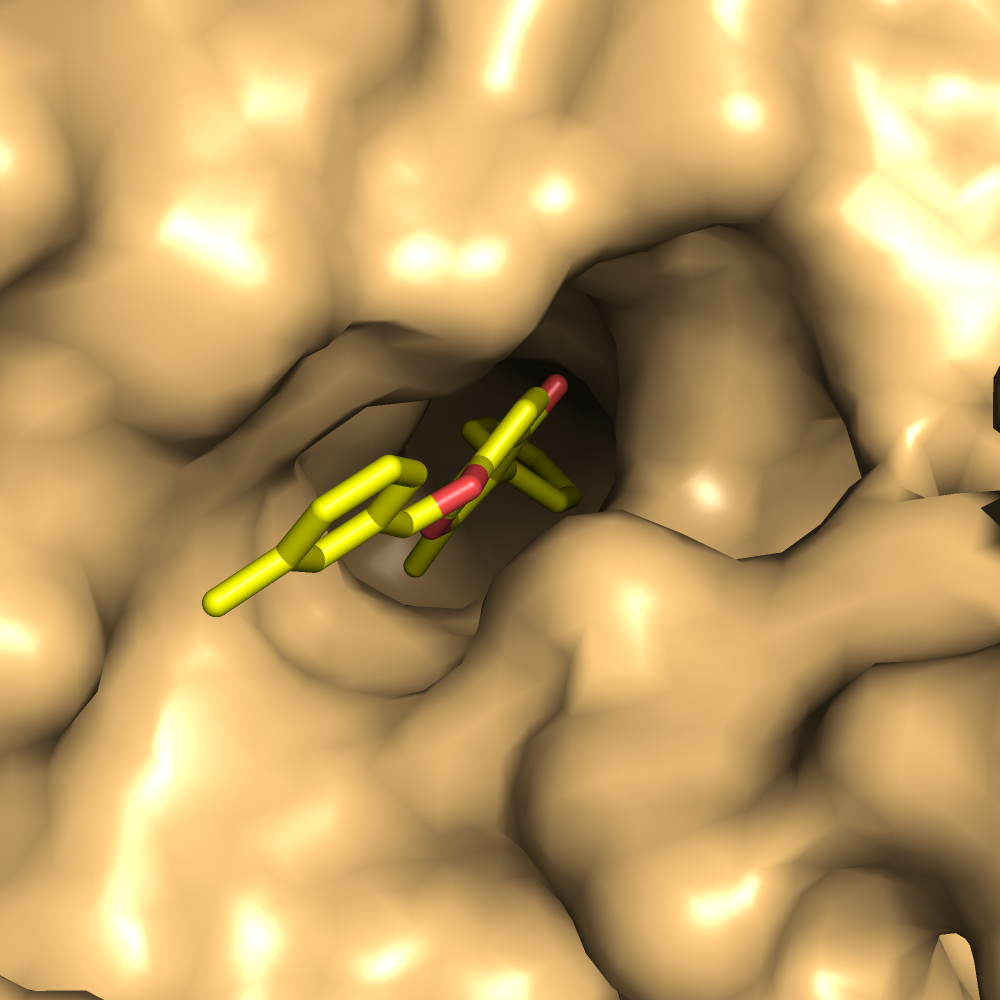} \\ \vspace{0.2cm}
    \includegraphics[width=\textwidth]{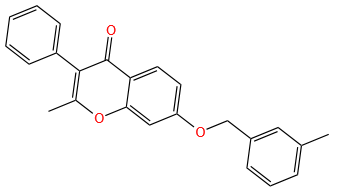}
    \centerline{Candidate 2}
\end{minipage}
\begin{minipage}{0.19\textwidth}
    \centering
    \includegraphics[width=\textwidth]{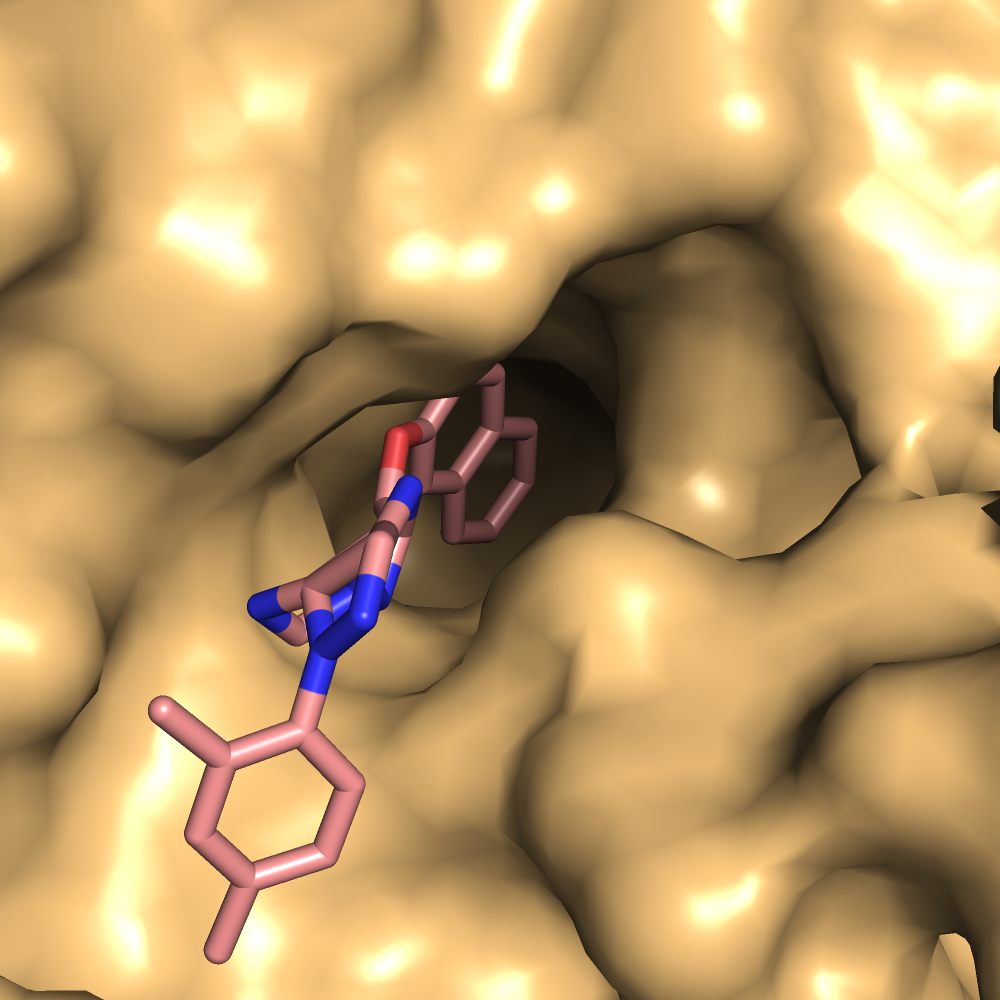} \\ \vspace{0.1cm}
    \includegraphics[width=\textwidth]{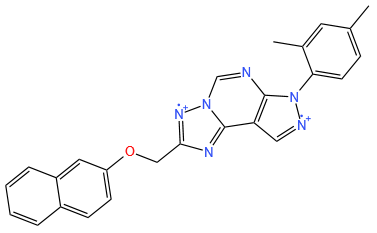}
    \centerline{Candidate 3}
\end{minipage}
\begin{minipage}{0.19\textwidth}
    \centering
    \includegraphics[width=\textwidth]{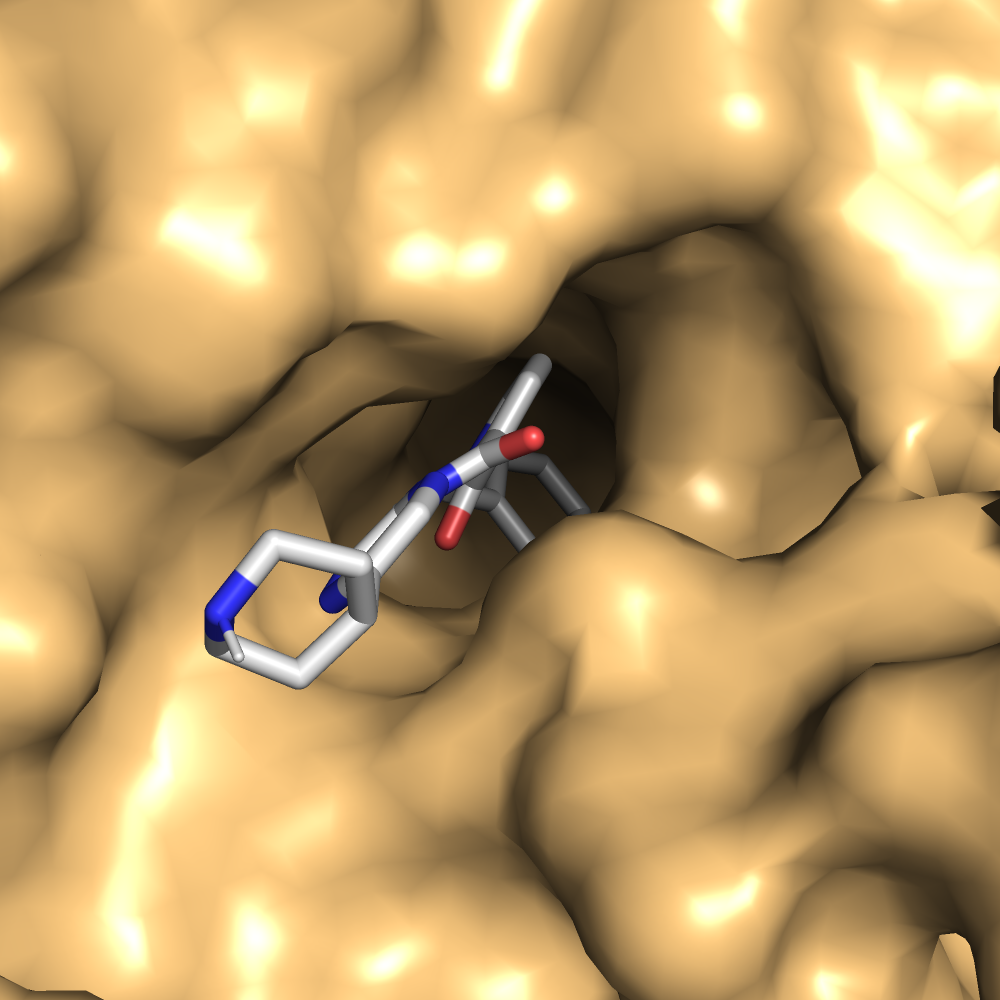} \\
    \includegraphics[width=\textwidth]{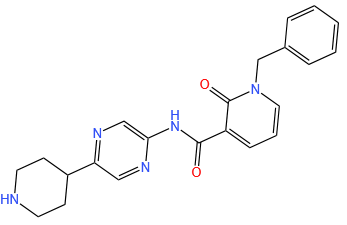}
    \centerline{Candidate 4}
\end{minipage}
\caption{Case study on protein target 1R1H: structures and binding poses of ligands.}
\label{1r1h}
\end{figure*}
\begin{figure*}[htb]
\centering
\begin{minipage}{0.19\textwidth}
    \centering
    \includegraphics[width=\textwidth]{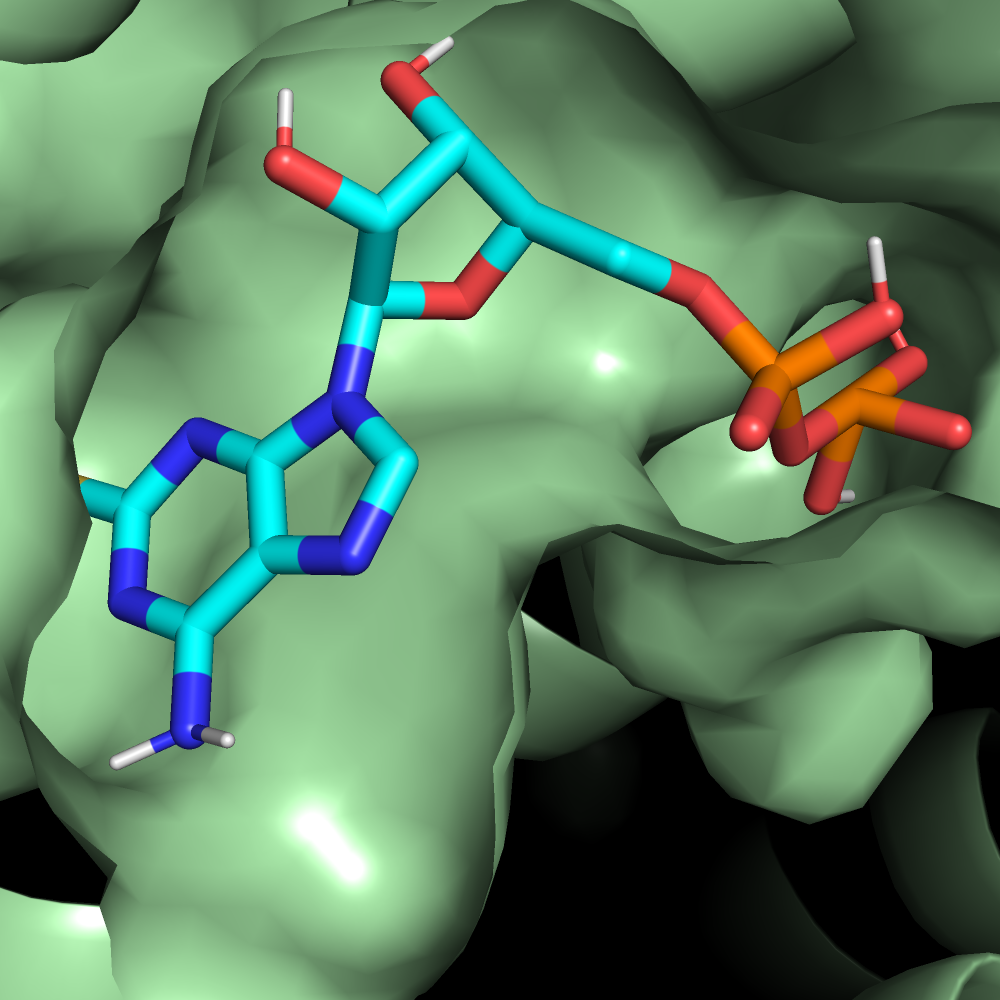} \\ \vspace{0cm}
    \includegraphics[width=\textwidth]{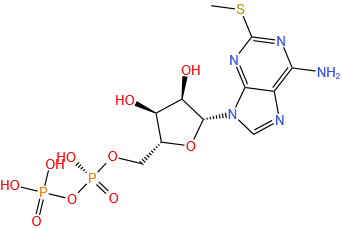}
    \centerline{Reference}
\end{minipage}
\begin{minipage}{0.19\textwidth}
    \centering
    \includegraphics[width=\textwidth]{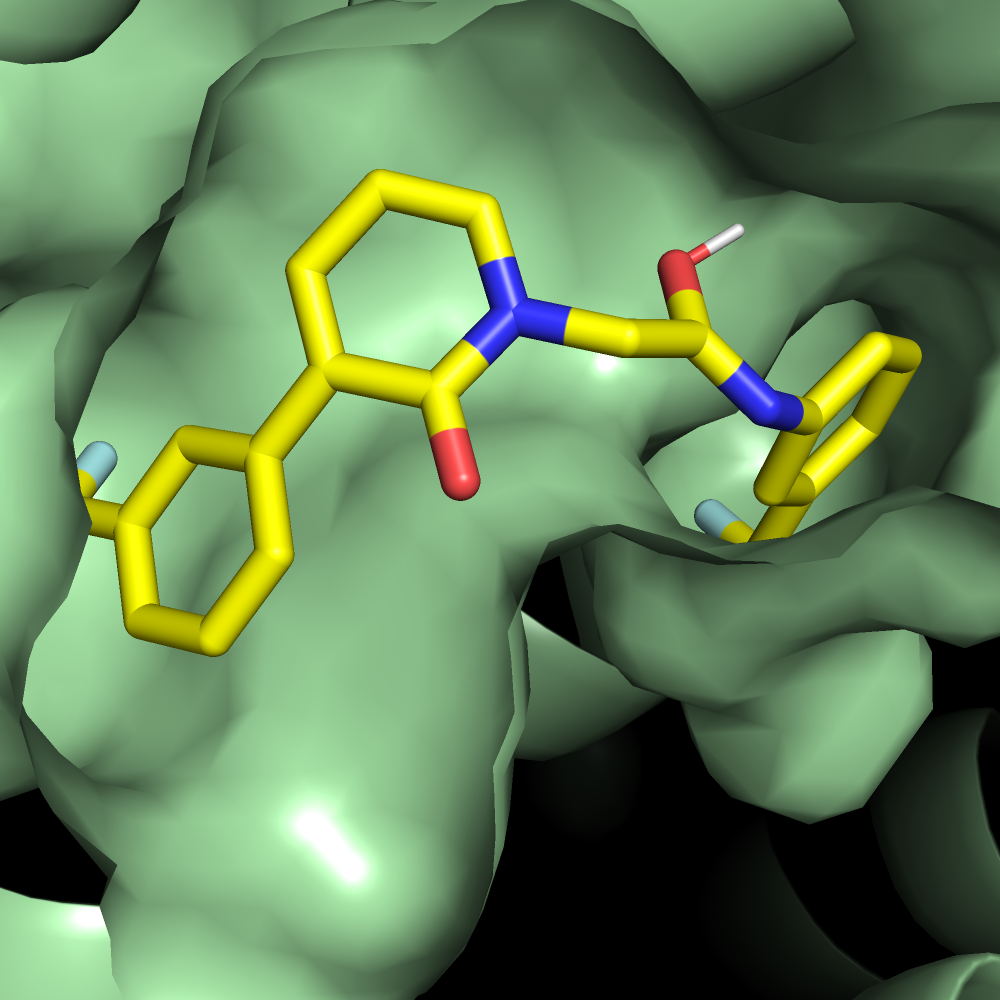} \\ \vspace{0.1cm}
    \includegraphics[width=\textwidth]{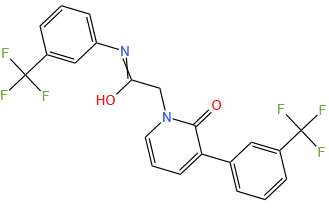}
    \centerline{Candidate 1}
\end{minipage}
\begin{minipage}{0.19\textwidth}
    \centering
    \includegraphics[width=\textwidth]{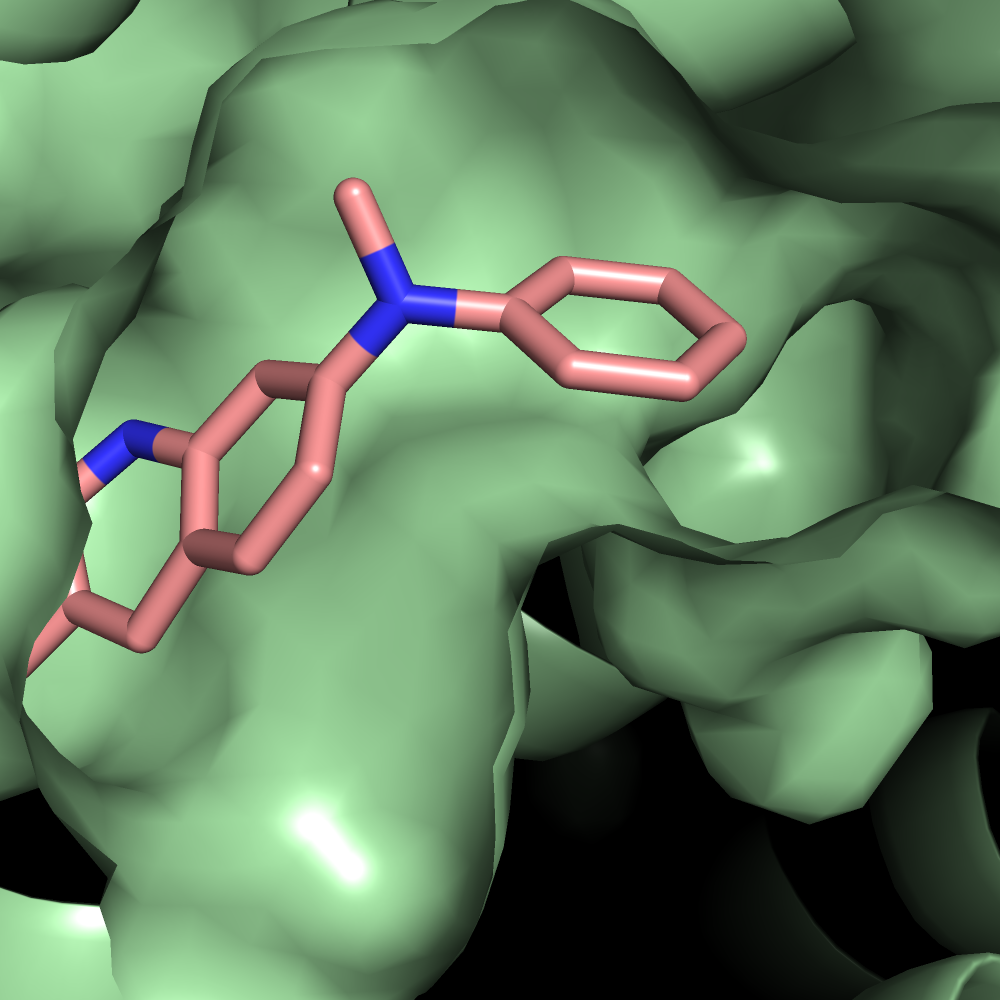} \\ \vspace{0cm}
    \includegraphics[width=\textwidth]{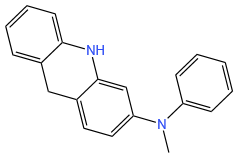}
    \centerline{Candidate 2}
\end{minipage}
\begin{minipage}{0.19\textwidth}
    \centering
    \includegraphics[width=\textwidth]{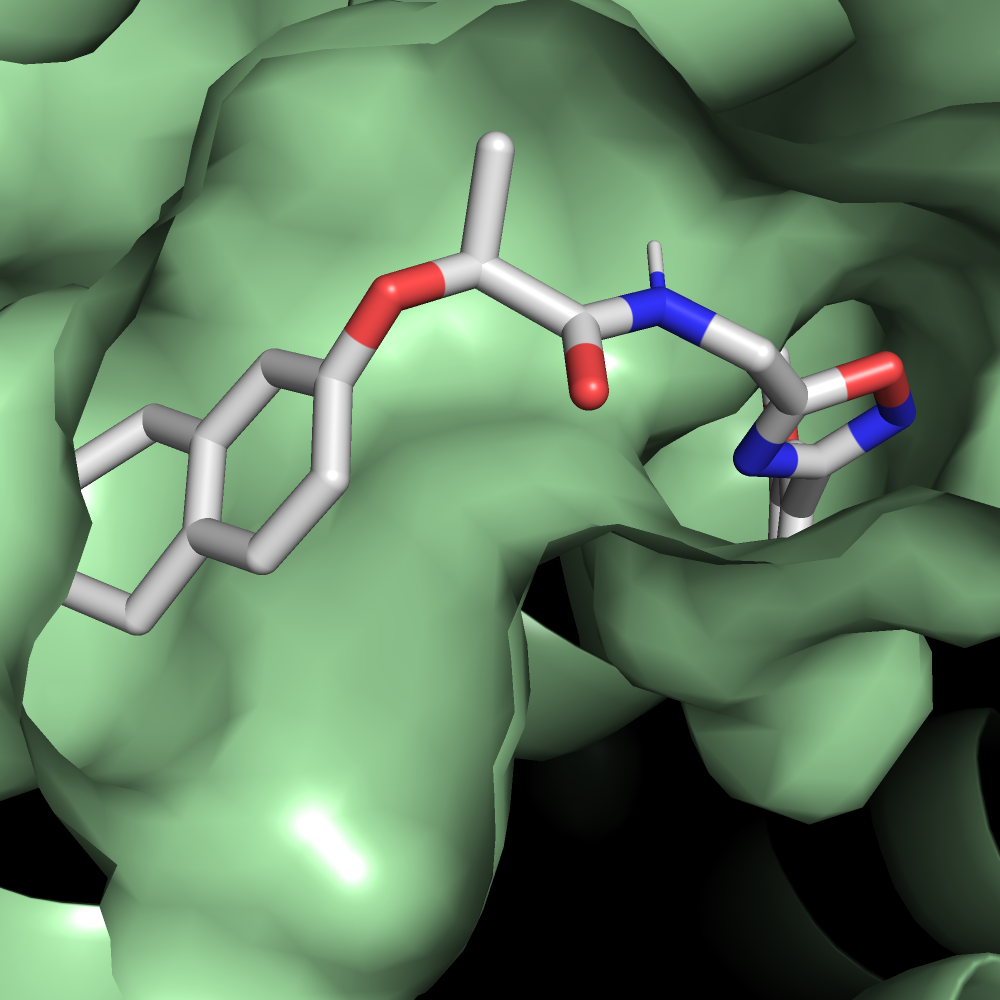} \\ \vspace{0.18cm}
    \includegraphics[width=\textwidth]{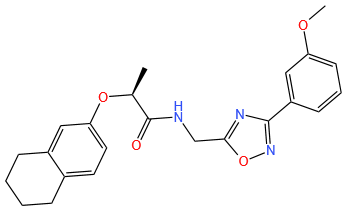}
    \centerline{Candidate 3}
\end{minipage}
\begin{minipage}{0.19\textwidth}
    \centering
    \includegraphics[width=\textwidth]{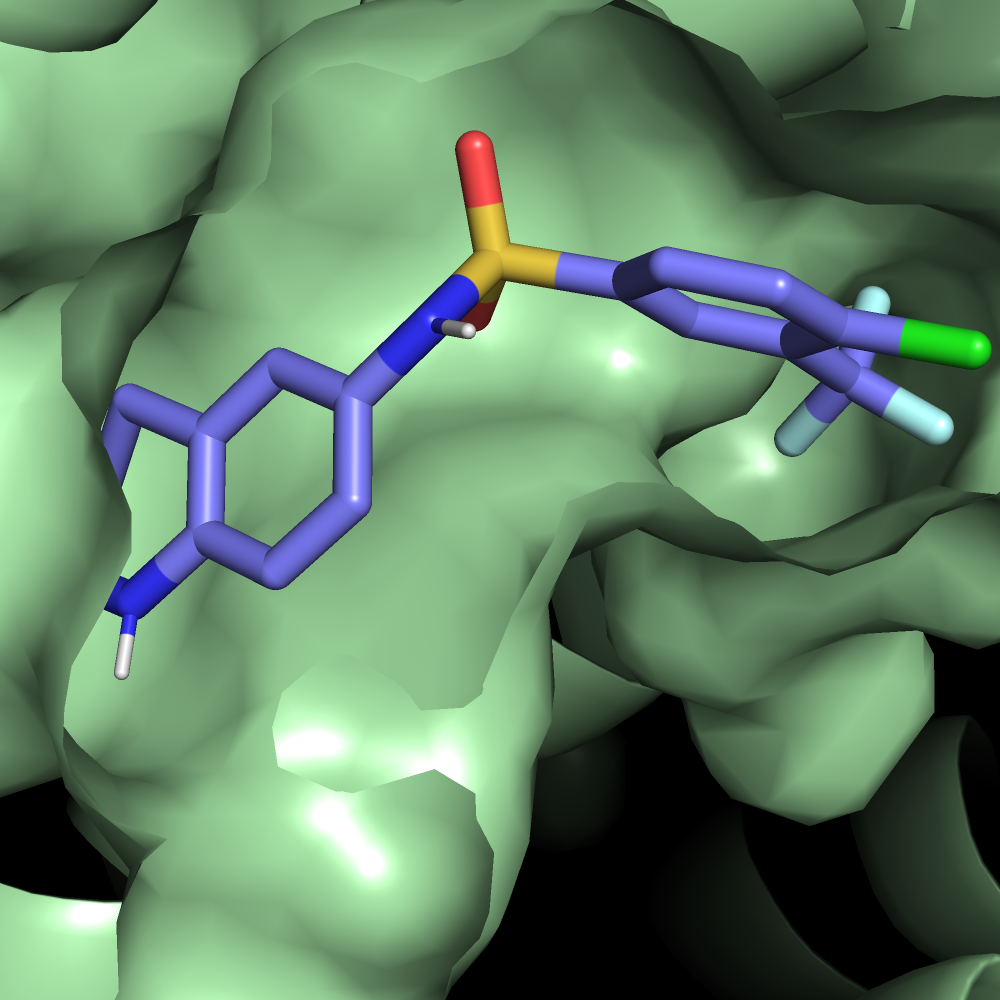} \\ \vspace{0.25cm}
    \includegraphics[width=\textwidth]{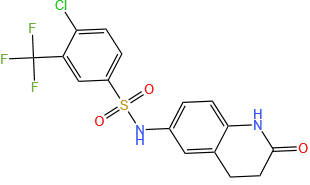}
    \centerline{Candidate 4}
\end{minipage}
\caption{Case study on protein target 4PXZ: structures and binding poses of ligands.}
\label{4pxz}
\end{figure*}

\subsection{Experimental Setup}
\paragraph{Dataset}
We conduct experiments on the widely used CrossDocked2020 dataset \cite{CrossDocked2020}, a commonly recognized benchmark for structure-based drug design. Following previous works \cite{Pocket2Mol,DecompDiff}, we adopt the same train-test split, with the training set containing 100,000 protein pocket-ligand complexes and the test set consisting of 100 protein pockets. The test set is filtered to include pockets with low sequence similarity to those in the training set. Additionally, the dataset of 3D small molecules collected by Uni-Mol \cite{Uni-Mol} is utilized for large-scale pretraining, which contains around 209M samples.

\paragraph{Baselines}
Our TransDiffSBDD model is compared against several recent baselines for SBDD, including AR \cite{AR}, liGAN \cite{liGAN}, GraphBP \cite{GraphBP}, Pocket2Mol \cite{Pocket2Mol}, TargetDiff \cite{TargetDiff}, FLAG \cite{FLAG}, DecompDiff \cite{DecompDiff}, DecompOpt \cite{DecompOpt}, and MolCRAFT \cite{MolCRAFT}. Additionally, since recent studies have demonstrated the competitiveness of 1D/2D molecular generation methods combined with docking software, we include two such methods, Reinvent \cite{Reinvent,Reinvent4} and RGA \cite{RGA}, as baselines. These represent state-of-the-art approaches for 1D and 2D molecular generation, respectively, and are both RL-based methods. For reference, we also include the ligands corresponding to the 100 pockets in the CrossDocked2020 test set.

\paragraph{Evaluation}
For evaluation, each method generates 100 candidate molecules for each protein pocket. Six metrics were used to comprehensively evaluate the practical value of these candidates in real-world drug discovery:
\begin{itemize}
    \item \textbf{Vina Score}: The binding affinity of the generated 3D molecules with the protein pockets, assessed using AutoDock Vina \cite{AutoDockVina} (units: kcal/mol).
    \item \textbf{Vina Dock}: The binding affinity of the generated molecules after re-docking with the protein pockets via AutoDock Vina (units: kcal/mol).
    \item \textbf{QED} (Quantitative Estimate of Drug-likeness) \cite{QED}: A desirability-based measure of drug-likeness, ranging from $[0,1]$.
    \item \textbf{SA} (Synthetic Accessibility) \cite{SA}: An estimate of the structural synthetic accessibility. The original score ranges from $[1,10]$ and is negatively linearly mapped to $[0,1]$, following common practice \cite{DESERT}.
    \item \textbf{Diversity} (Internal Diversity) \cite{IntDiv}: The average of pairwise Tanimoto distances between extended-connectivity fingerprints \cite{ECFP} of a set of molecules, ranging from $[0,1]$.
    \item \textbf{Success Rate}: The percentage of generated molecules meeting the criteria: Vina Dock $<-8.18$ kcal/mol, QED $>0.25$, and SA $>0.59$, as introduced by \citet{DESERT}, which simulates the practical demand of multi-property optimization in drug discovery.
\end{itemize}
For each metric, the average value of the generated molecules across 100 protein pockets is reported.

\subsection{Results}
As shown in Table \ref{results}, our TransDiffSBDD outperforms all baselines in Vina Dock, SA, and Success Rate, while also achieving the second-best performance in Vina Score and Diversity, and maintaining competitive results in QED. These results indicate that TransDiffSBDD provides significant advantages over existing methods for structure-based drug design. In particular, its superior Success Rate highlights its potential for real-world drug discovery scenarios where multi-objective optimization is crucial.

Additionally, we observe that MolCRAFT performs well in Vina Score and Vina Dock, both of which directly measure binding affinity, while Reinvent+Vina achieves higher scores in Diversity and Success Rate, metrics more aligned with practical drug discovery needs. MolCRAFT formulates the SBDD task entirely in continuous parameter space, representing discrete atom types using a categorical distribution. This approach facilitates continuous 3D modeling but compromises the discrete nature of molecules and neglects the causal relationship between discrete and continuous information. On the other hand, Reinvent+Vina, while lacking explicit 3D molecular modeling, effectively captures discrete molecular structures in an autoregressive manner and benefits from pretraining on large-scale small molecule datasets and flexible RL objectives.
Notably, TransDiffSBDD combines the strengths of both approaches, consequently achieving comprehensive superiority over existing methods. These results underscore the effectiveness of our causal-aware multi-modal modeling approach for structure-based drug design.

\subsection{Case Studies}
\begin{table}[tb]
\centering
\caption{Case studies on targets 1R1H and 4PXZ: property scores of the ligands shown in Figures \ref{1r1h} and \ref{4pxz}. Ref. and Cand. are short for Reference and Candidate, respectively.}
\label{casestudy}
\small
\begin{tabular}{ccccc}
\hline
Target & Ligand & Vina Dock ($\downarrow$) & QED ($\uparrow$) & SA ($\uparrow$) \\ \hline
\multirow{4}{*}{1R1H} & Ref. & -7.6 & 0.46 & 0.73\\
& Cand. 1 & -10.5 & 0.34 & 0.85\\
& Cand. 2 & -10.4 & 0.47 & 0.89\\
& Cand. 3 & -10.3 & 0.65 & 0.68\\
& Cand. 4 & -10.2 & 0.70 & 0.84\\ \hline
\multirow{4}{*}{4PXZ} & Ref. & -6.6 & 0.16 & 0.64\\
& Cand. 1 & -10.9 & 0.31 & 0.82\\
& Cand. 2 & -10.5 & 0.55 & 0.87\\
& Cand. 3 & -10.2 & 0.64 & 0.81\\
& Cand. 4 & -9.9 & 0.81 & 0.85\\ \hline
\end{tabular}
\end{table}

Beyond large-scale benchmark experiments, we conduct case studies on inhibitor design for two protein targets from the CrossDocked2020 test set: 
(1) Target \textbf{1R1H} \cite{1R1Htarget}: Neutral endopeptidase (NEP), which plays a key physiological role in modulating human nociceptive and pressor responses.
(2) Target \textbf{4PXZ} \cite{4PXZtarget}: P2Y12 receptor (P2Y12R), a prominent clinical drug target for the inhibition of platelet aggregation. 
For each target, we present the reference ligand from CrossDocked2020 and four candidate ligands generated by TransDiffSBDD.

As shown in Figures \ref{1r1h}, \ref{4pxz} and Table \ref{casestudy}, neither of the reference ligands meets the "success" criteria. In contrast, all candidate molecules designed by TransDiffSBDD satisfy the criteria and exhibit much better Vina Dock scores compared to the reference ligands. Furthermore, for each target, TransDiffSBDD generates ligands with diverse structures and binding mechanisms, further demonstrating the practical potential of our approach in drug discovery.
\section{Conclusion and Discussion}
In this work, we identify two key limitations of existing methods for structure-based drug design: (1) the inability to model the modality differences between a ligand’s discrete graph information and continuous 3D information, and (2) the failure to account for the causality between these two modalities. To address these issues, we propose TransDiffSBDD, a causality-aware multi-modal framework for SBDD, which integrates an autoregressive transformer and a diffusion model to effectively handle modality differences. Furthermore, we design hybrid-modal sequences that explicitly preserve the causality between molecular representations, enabling a more principled formulation of the SBDD task. Experimental results demonstrate the robustness and practical competitiveness of our approach across multiple evaluation metrics.

Nevertheless, there is still room for improvement. First, our model incorporates the stochasticity of atomic coordinates from the diffusion process, which allows for modeling the equilibrium distribution of molecular 3D structures. However, due to data scarcity, we are currently unable to fully realize this potential. Second, protein-ligand binding is inherently a dynamic temporal process, but like other existing methods, our approach is limited by the lack of time-resolved binding datasets to model these dynamics effectively. Finally, our model lacks direct interpretability, which may hinder its adoption by biopharmaceutical researchers in practical drug discovery applications. These limitations highlight important directions for future research on SBDD.

\section*{Impact Statement}
This paper focuses on the application of machine learning in drug discovery, which has the potential to advance pharmaceutical research and contribute to human health. The algorithm itself does not cause any direct social consequences, however, it may serve as a tool for drug development. Therefore, its usage should be regulated to prevent the design and production of toxic, harmful molecules or other unlawful activities.

\newpage

\bibliography{ref}
\bibliographystyle{icml2025}

\end{document}